\documentclass[reprint]{revtex4-1}

\usepackage{graphicx} 
\usepackage{amsmath}
\usepackage{ amssymb }

\usepackage{float} 
\usepackage{wrapfig} 

\usepackage[makeroom]{cancel} 
\usepackage{comment}

\usepackage{lipsum} 

\newcommand{\beq}{\begin{equation}}
\newcommand{\eeq}{\end{equation}}
\newcommand{\vv}{\mathbf{v}}

\newcommand{\bvec}{\begin{pmatrix}}
\newcommand{\evec}{\end{pmatrix}}
\newcommand{\lp}{\left(}
\newcommand{\rp}{\right)}

\newcommand{\eps}{\epsilon}

\newcommand{\pa}[2]{\frac{\partial #1}{\partial #2}}
\newcommand{\tot}[2]{\frac{d #1}{d #2}}

\newcommand{\Rv}{\mathbf{R}}
\newcommand{\Ev}{\mathbf{E}}
\newcommand{\Bv}{\mathbf{B}}

\newcommand{\tn}{\tilde{ n}}
\newcommand{\tr}{\tilde{ r}}
\renewcommand{\tt}{\tilde{ t}}
\newcommand{\tT}{\tilde{ T}}
\newcommand{\tR}{\tilde{ R}}

\begin{document}



\title{Favorable Collisional Demixing of Ash and Fuel in Magnetized Inertial Fusion}

\author{Ian E. Ochs and Nathaniel J. Fisch}
\affiliation{Department of Astrophysical Sciences, Princeton University, Princeton, New Jersey 08540, USA}


\date{\today}

\begin{abstract}
	Magnetized inertial fusion experiments are approaching regimes where the radial transport is dominated by collisions between magnetized ions, providing an opportunity to exploit effects usually associated with steady-state magnetic fusion.
	In particular, the low-density hotspot characteristic of magnetized liner inertial fusion 
	results in diamagnetic and thermal frictions which can demix thermalized ash from fuel, accelerating the fusion reaction.
	For reactor regimes in which there is substantial burnup of the fuel,   increases in the fusion energy yield on the order of 5\% are possible.
	\begin{description}
		\item[PACS numbers]
	\end{description}
\end{abstract}

\maketitle





\textbf{Introduction:}  In inertial deuterium-tritium (DT) fusion, stratification of different ion species can significantly impact the fusion energy output.
Demixing of the fuel ions  reduces the fusion reaction rate \cite{kagan2012electro,kagan2014thermodiffusion,bellei2013species, keenan2018ion, vold2018self, vold2017plasma}.
Mixing of the fuel   with impurities or, in the case of large  burnup, with fusion ash,  also reduces the fusion reaction.
Ideally, the ash and impurities should be separated from the fuel, while the fuel itself should remain as mixed as possible.


Importantly, ion stratification  in magnetized inertial fusion (MIF) can enter magnetized transport regimes more traditionally associated with steady-state magnetic fusion energy (sMFE) \cite{braginskii1965transport, hinton1983collisional}.
In sMFE, where the density is peaked on-axis,  diamagnetic frictions  drive high-$Z_I$ impurities into the high-density core region on the ion-ion diffusion timescale \cite{spitzer1952equations, taylor1961diffusion}.
To mitigate this deleterious effect, the temperature is highly-peaked on-axis, which tends to flush impurities outward as a result of  thermal friction \cite{ hirshman1981neoclassical, helander2005collisional,valisa2011metal,lerche2016optimization}.
In contrast, in MIF devices such as Magnetized Liner Inertial Fusion (MagLIF), the density is naturally peaked at the plasma edge, while the temperature  is naturally peaked at the core \cite{slutz2010pulsed, mcbride2015semi}; thus both the density and temperature profiles are naturally arranged to transport fusion ash ($\alpha$ particles) outwards.

In fact, as we show here, the MIF stagnation time can be comparable to the ion-ion diffusion time, yet  much shorter than the ion-electron diffusion timescale on which the effects dissipate.  Thus in the event of a large fuel burnup fraction, with a corresponding copious production of ash, there is also a mechanism for ash expulsion, which purifies the plasma.
Large burnup fractions would be necessary for any economical implementation  of MIF as an energy source.  
Moreover, the collisional mechanism that expels the α particles also naturally replenishes the fuel ions.
This can lead, as we show here, to significant enhancements in the burnup fraction, on the order of 5\% for burnup fractions as low as 25\%. 
The model that we offer for the  compression  is highly idealized  and simplified, but it does illustrate the significant opportunities.

%

\begin{figure}[b]
	\center
	\includegraphics[width=\linewidth]{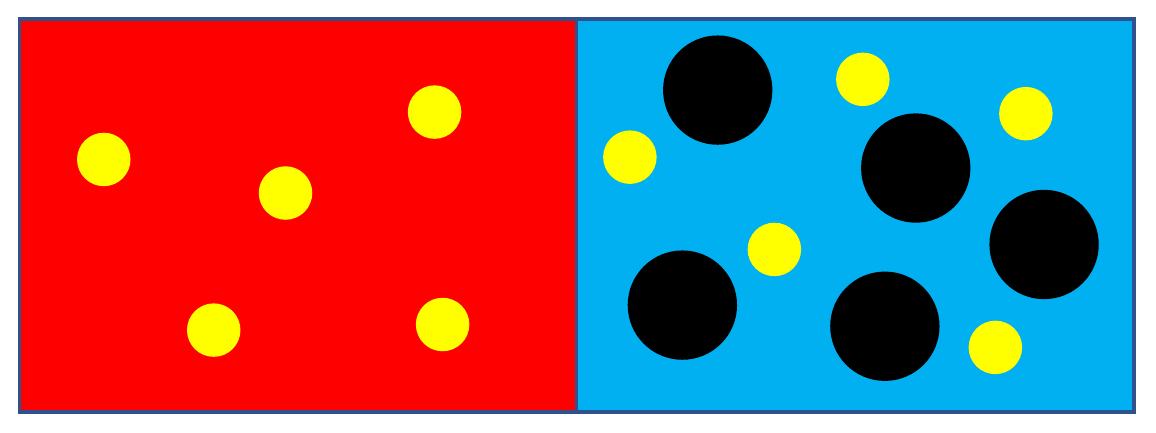}
	\caption{
		Advantage of demixing. 
		Suppose $N = 10$ yellow fuel ions, and $M = 5$ black ash ions with $Z = 2$. 
		Box 1 (left) contains $1/4$ of the total charge, with $T_1 / T_2 = 2$.
		Fusion power is greatest when all the ash ions are in the colder box.
	}
	\label{fig:optimization} 
\end{figure}

The favorable expulsion occurs because ion-ion collisions  rearrange magnetized  fuel and ash without moving net charge.
Thus, fuel and ash can be exchanged across the magnetic field, as long as the local charge remains unchanged.
Because the fusion reaction scales strongly with temperature, reactivity is maximized with  fuel  concentrated in the hot, rarified regions, and ash in the cold, dense regions (Fig.~\ref{fig:optimization}).
Fortuitously,  this is exactly the result of classical magnetized transport.

\textbf{Model:} Denote the impurity (ash) by the subscript $I$, and the hydrogenic fuel ions by the subscript $H$.
In a magnetized plasma, pressure and temperature gradients will give rise to azimuthal diamagnetic drifts which differ for each species.
This will in turn give rise to azimuthal diamagnetic and thermal frictions \cite{braginskii1965transport}, which result in radial $F \times B$ drifts that constitute the diffusive transport.
For compression times long compared to an impurity-ion collision time $\tau_{IH}$, and a compression velocity slow compared to the thermal velocity $\sqrt{T/m_i}$, the collisional transport motion due to ion-ion collisions is given by \cite{hinton1983collisional}:
\begin{align}
	\vv_I^{(Tr)} &= -\frac{T}{m_I \Omega_I^2 \tau_{IH}} \biggl\{  \frac{\nabla_\perp n_I }{n_I}  - Z_I \frac{\nabla_\perp n_H  }{n_H} \notag \\
	& \quad + \left[ 1 + \lp \frac{3}{2} H_{HI} - 1 \rp Z_I \right]  \frac{\nabla_\perp T}{T} \biggr\}, \label{eq:diff}
\end{align}
where $\Omega_I$ is the impurity gyrofrequency, and 
\beq
H_{HI} = \frac{1}{\sum_H n_H} \sum_H n_H \lp \frac{1 - Z_H m_H / Z_I m_I}{1 + m_H/m_I} \rp
\eeq
is a numerical factor that determines the strength of the thermal friction, averaged over all hydrogenic species present \cite{hinton1983collisional}.
For alpha ash interacting with an equal mix of deuterium and tritium, $H_{HI} = 3/7$.

To find the stationary state of the transport process,
take $\vv_I^{(Tr)} = 0$ and integrate Eq.~(\ref{eq:diff}) over space.
The steady state radial distribution obeys:
\begin{align}
n_I(r)  &\propto n_{H}(r)^{\frac{Z_I}{Z_H}} T(r)^{-\lp \frac{3}{2} H_{HI} - 1 \rp \frac{Z_I}{Z_H} -1}\label{eq:Hpinch}. 
\end{align}
Taking the infinite-mass-ratio limit $H_{HI} \rightarrow 1$ of Eq.~(\ref{eq:Hpinch}) yields the well-known classical impurity pinch result \cite{spitzer1952equations, taylor1961diffusion, hirshman1981neoclassical,helander2005collisional}. 
For $\alpha$ ash, with $H_{HI} = 3/7$, Eq.~(\ref{eq:Hpinch}) implies:
\begin{align}
	n_I(r)/n_{H}(r) &\propto n_{H}(r) T(r)^{-2/7} .\label{eq:alphaPinch} 
\end{align}
Thus we find the fortuitous result that the ash will tend to be relatively concentrated in regions of high density and low temperature.
The thermal friction is critical to this result; if we took the thermal-friction-free limit $H_{HI} \rightarrow 0$, the temperature dependence would invert, and Eq.~(\ref{eq:alphaPinch}) would become $n_I / n_H \propto n_H T \approx P$; i.e. there would be no demixing in an isobaric plasma.

\begin{figure*}
	\center
	\includegraphics[width=\linewidth]{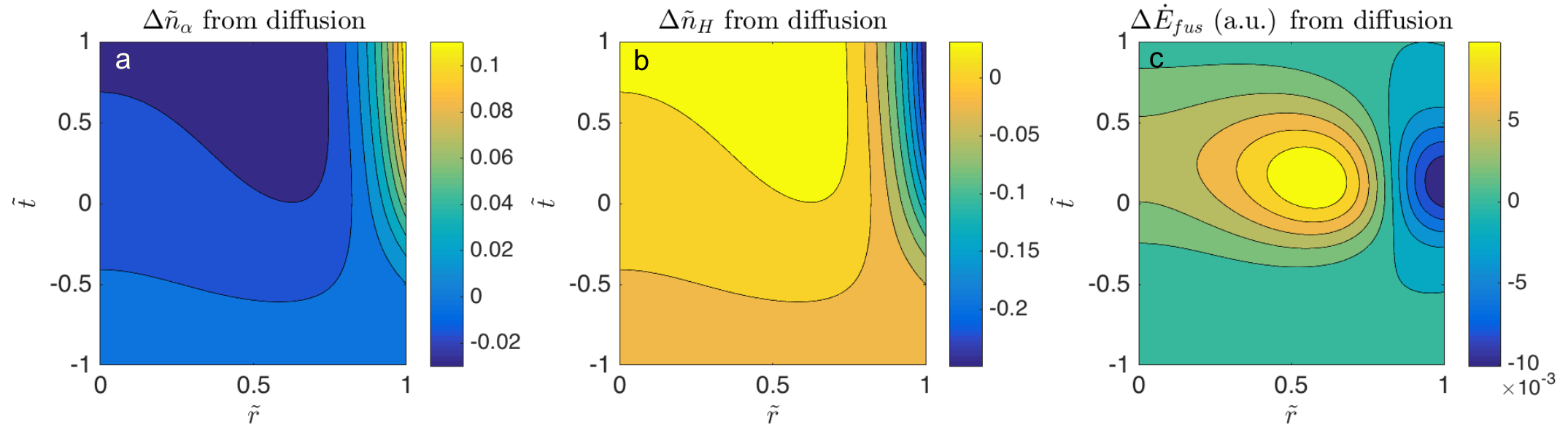}
	\caption{
		Change in parameters due to ion diffusion in a stagnating plasma with a hot central core   as a function of normalized radius and time:
		(a) $\alpha$ density; (b) fuel density;  (c) fusion reaction rate.
		Magnetized diffusion results in more fuel in the hot central region and less in the periphery,
		increasing the burn fraction by 4.8\% compared to a simulation that neglects  diffusion.
	}
	\label{fig:relFusion} 
\end{figure*}

To quantify the effect of this demixing on the energy yield of the fusion reaction, we combine our transport model with a model for the D-T fusion reaction:
\beq
	\pa{n_\alpha}{t}|_{\text{fus}} = \langle \sigma v \rangle(T) n_{D} n_{T} = \frac{1}{4} \langle \sigma v \rangle(T) n_{H}^2, \label{eq:fus}
\eeq
where we have taken $n_D = n_T = n_H / 2$.
For $200$ eV$ < T < 100$ keV, the reactivity $\langle \sigma v \rangle (T)$ can be analytically approximated to within $0.25\%$ accuracy \cite{bosch1992improved}.

Consider self-similar compression \cite{velikovich1985hydrodynamics,ramsey2010class}, where the radial velocity due to compression is given by $v_r^{(C)}(r,t) = r \dot{R}(t)/R(t)$.
This ensures that for a fluid element, $\tr \equiv r(t) / R(t)$ is constant in time.
The function $R(t)$ will evolve on some characteristic stagnation timescale $\tau_{s} \sim R(t) / \dot{R}(t)$.
The variables $\tr$ and $\tt \equiv t/\tau_{s}$ thus represent the ``natural'' independent variables of the problem, with associated normalized dependent variables $\tn_H$, $\tn_I$, $\tT$, and $\tR$, related to the dimensional variables by
\begin{align}
	\tR(\tt) &= R(t) / a\\
	n_H(r, t) &= \tn_H(\tr, \tt) n_{H0a} \tR(\tt)^{-2} \label{eq:nEvo1}\\
	n_I(r, t) &= \tn_I(\tr, \tt) n_{H0a} \tR(\tt)^{-2} \label{eq:nEvo2}\\
	T (r, t) &= \tT(\tr, \tt) T_{0a} \tR(\tt)^{-4/3} \label{eq:TEvo},
\end{align} 
where $a$ is the stagnation radius and $n_{H0a}$ and $T_{0a}$ are typical densities and pressures at the plasma edge at stagnation.
Eqs.~(\ref{eq:nEvo1}-\ref{eq:nEvo2}) follow from the fuel and ash continuity equations, while Eq.~(\ref{eq:TEvo}) follows from adiabatic compression of the cylindrical plasma.
We assume that the point of maximal compression occurs at $\tt = 0$, so that $\tR(\tt=0) = 1$, with $\tR(\tt) \geq 1$.
Our explicit normalizations on $\tn$ and $\tT$ follow from the fact that we enforce that $\tT=1$ and $\tn_H=1$ at $(\tt=\tt_0,\tr=1)$, where $\tt_0$ is the initial simulation timepoint, satisfying $\tt_0 \leq 0$.
Thus $n_{H0a}$ and $T_{0a}$ represent the values of $n_H$ and $T$ at $(t = 0,r=a)$ in the absence of any diffusion or fusion, i.e. if the initial conditions were simply self-similarly compressed.
Note that $\tn_I$ is normalized to the value of $\tn_H$ at $\tr=1$, so that the relative densities of different species can be compared.

The impurities will obey a continuity equation, incorporating the velocities due to both transport (Eq.~(\ref{eq:diff})) and compression, as well as a source term arising from the fusion reaction (Eq.~(\ref{eq:fus})), giving
\beq
\pa{n_I}{t} + n_I \nabla \cdot \vv_I = \frac{1}{4} n_H^2 \langle \sigma v \rangle (T).
\eeq 
Plugging our velocities into this continuity equation and transforming into our nondimensionalized variables, we find the governing equation for the ash distribution:
\begin{align}
\pa{\tn_I}{\tt} &= \frac{1}{\tr} \pa{}{\tr} \biggl\{ \tr \tn_I D_0(\tr, \tt) \biggl[ 
\frac{1}{\tn_I} \pa{\tn_I}{\tr} - Z_I \frac{1}{\tn_H} \pa{\tn_H}{\tr} \notag\\
&\qquad \qquad + \lp 1 + \lp \frac{3}{2} H_{HI} - 1 \rp Z_I \rp \frac{1}{\tT} \pa{\tT}{\tr}  \biggr]  \biggr\} \notag\\
&\qquad + S(\tt) \langle \sigma v \rangle (\tT T_{0a} \tR(\tt)^{-4/3}) \tn_H^2, \label{eq:diffModel}
\end{align}
where ambipolar transport, constant pressure, and magnetic flux compression imply that
\begin{align}
\tn_H &= \tn_{H0} + Z_I (\tn_{I0} - \tn_I) \label{eq:ambipolar}\\
\tT &= \lp \tn_I + \tn_H \rp^{-1}, \label{eq:constP}\\
D_0(\tr, \tt) &= D_{0a} \tn_H \tT^{-1/2} \tR(\tt)^{2/3}\\
S(\tt) &= S_{0a} \tR(\tt)^{-2}, \label{eq:S}
\end{align}
where $S_{0a} = n_{H0a} \tau_{s} / 4$.

The relevant dynamical timescales are determined by the initial steepness of the distribution, as well as two dimensionless constants.
The first constant is the number of diffusion times in a typical stagnation time $\tau_{s}$:
\begin{align}
	D_{0a} &= \lp \frac{\rho_{H0a}}{a} \rp^2 \frac{\tau_{s}}{\tau_{IH0a}} \label{eq:D0a}\\
	 & \approx 0.7 n_{23}  T_{10}^{-1/2} B_{10}^{-2}  a_{10}^{-2} \tau_{s1},
\end{align}
where $\rho_{H0a}$ is the fuel gyroradius, $n_{23} = n_{0a} / (10^{23}$ cm$^{-3})$, $T_{10} = T_{0a} / (10$ keV), $B_{10} = B/$(10 kT), $a_{10} = a/$(10 $\mu$m), and $\tau_{s1} = \tau_{s} /$(1 ns).
The second constant is the number of fusion burn times in a stagnation time:
\begin{align}
	S_{0a} \langle \sigma v \rangle_{0a} &= \frac{n_{H0a}\tau_{s}}{4} \langle \sigma v \rangle (T_{0a})\\
	&\approx 0.003 n_{23} \tau_{s1} T_{10}^2 \text{ for } 0.5<T_{10} < 2.
\end{align}
For a centrally-peaked temperature, the diffusion coefficient will tend to decrease from $D_{0a}$ toward the center of the plasma, while the fusion coefficient will tend to increase toward the center.
There will be significant effects due to both diffusion and fusion over a stagnation time if the typical diffusion coefficient in the plasma is $\gtrsim 1$, and the typical fusion coefficient is $\lesssim 1$.
Meanwhile, our magnetized transport assumption will be valid as long as the gyroradius is much smaller than the system size, i.e. $\rho_I / a \ll 1$, and the collision frequency is much smaller than the gyrofrequency, i.e.
\begin{align}
	\Omega_I \tau_{IH} & = 7 n_{23}^{-1} B_{10} T_{10}^{3/2}  \gg 1.\label{eq:magnetization}
\end{align}

\textbf{Stagnation scenario:} 
To describe a MagLIF stagnation, we take a form of $\tR(\tt)$ suggestive of compression, followed by stagnation for a characteristic time $\Delta \tt_s = 1$, followed by expansion:
\beq
	\tR(\tt) = 1 + \tt^2, \quad -1 < \tt < 1.
\eeq 	
To simulate the low-density hotspot, we take:
\begin{align}
\tn_{H0} = e^{\alpha (\tr^2 - 1) } \quad \tT_0 = e^{\alpha (1-\tr^2)}, \label{eq:Tprof}
\end{align}
where the steepness parameter $\alpha = \log(T_{0h} / T_{0a})$, with $T_{0h}$ the hotspot temperature at maximum compression. 

Consider a MagLIF-like stagnation scenario, with $B_0 = 25$ kT, $T_{0a} = 8$ keV, $n_{0a} = 7 \times  10^{23}$ cm$^{-3}$, $\tau_s = 8 \text{ ns}$, $a = 30$ $\mu$m.
We take a steepness parameter $\alpha = 1.5$, corresponding to core density and temperature $n_{0h} = 1.6 \times 10^{23}$ and $T_{0h} = 36$ keV.
For this choice of parameters, $D_{0a} = 0.68$ and $S_{0a} \langle \sigma v \rangle_{0a} = 0.087$, while at the core $D_{0h} = 0.1$ and $S_{0h} \langle \sigma v \rangle_{0h} = 0.24$.
This is on the slower and higher-field end for MagLIF stagnation parameters \cite{ryutov2015characterizing}, which we adopt to make sure that our orderings $\Omega_I \tau_{IH} > 1$ and $\rho_I / a  \ll 1$ remain valid throughout most of the plasma for the duration of the simulation; at $t=0, r=a$, we have $\Omega_I \tau_{IH}= 2.1$ and $\rho_I / a = 0.012$.
The brief periods at the beginning and end during which the edge plasma is unmagnetized should not have a substantial impact on the results, since (a) the majority of the plasma is more magnetized than the edge, and (b) most reactions occur near maximal compression, when the edge plasma is magnetized.

Simulations are carried out both with and without the diffusion terms.
The diffusion effects pull $\alpha$  ash out of the fusion hotspot to lower temperature (Fig.~\ref{fig:relFusion}a).
Because of the ambipolar transport constraint, this exclusion of ash increases the fuel concentration in the hotspot (Fig.~\ref{fig:relFusion}b).
This in turn increases the fusion power in the hotspot, and decreases the fusion power at the edge (Fig.~\ref{fig:relFusion}c).
%

The burnup fraction can be put as
\beq
F_{burn} = \frac{\int_0^1 (\tn_{H0} - \tn_H) \tr d\tr}{\int_0^1 \tn_{H0} \tr d\tr}. \label{eq:burnFrac}
\eeq
The impact of the diffusion terms can be captured by the fractional difference in burnup when diffusion is ignored or implemented,  
termed the {\it diffusive enhancement}:
\beq
G_D \equiv \frac{F_{burn}(\text{diffusion}) - F_{burn}(\text{no diffusion})}{F_{burn}(\text{no diffusion})}.
\eeq
The burn fraction $F_{burn}$ increases from 25.8\% to 27.0\% as a result of the demixing, leading to an enhancement of $G_D = 4.8\%$ in the neutron yield.
In a second example, similar enhancements are obtained for field-reversed magnetic target fusion parameters, i.e. $B_0 = 500$ T, $T_{0a} = 6$ keV, $n_{0a} = 1 \times  10^{20}$ cm$^{-3}$, $\tau_s = 50$ $\mu$s, $a = 1$ mm \cite{intrator2004high,intrator2004high2}, where $D_{0a} = 2.4$ and $S_{0a} \langle \sigma v \rangle_{0a} = 0.032$.



\begin{figure}[t]
	\center
	\includegraphics[width=\linewidth]{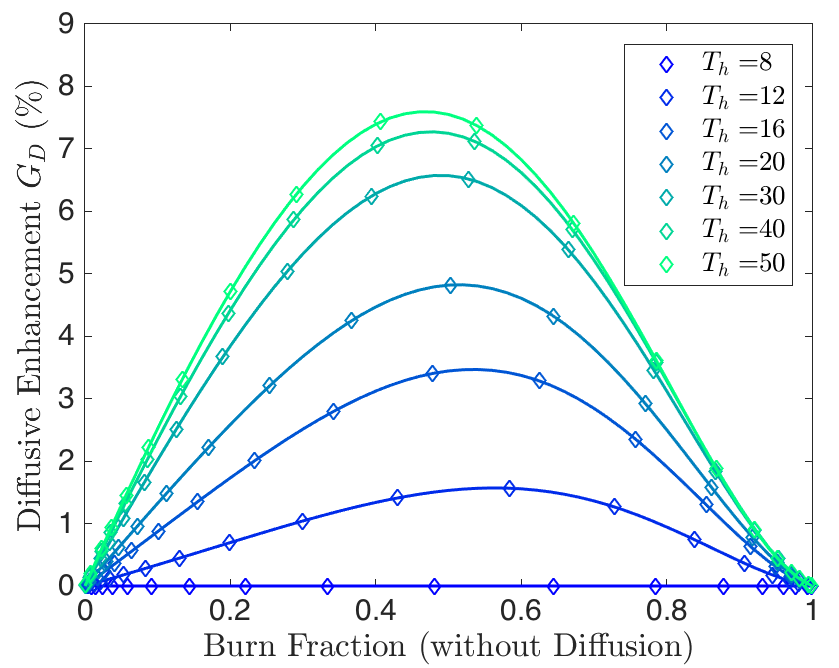}
	\caption{
		Total diffusive enhancement $G_D$ vs burn fraction $F_{burn} $(no diffusion) 
		for several values of the profile steepness $\alpha$ (Eqs.~\ref{eq:Tprof}).
		Points are simulation values; lines are spline interpolations.
		Higher enhancements occur at steeper density gradients, saturating at around 8\% at $T_{0h} = 50$.
	}
	\label{fig:burnScanAlpha} 
\end{figure}

\textbf{Temperature distribution effects:} 
The largest diffusive enhancements $G_D$ occur when the burnup fraction is large enough to poison the reaction, but not so large that the marginal effect of diffusion is insignificant; i.e. $G_D$ will be maximized in the regime around $F_{burn} \approx 50\%$, across a wide range of parameters (Fig.~\ref{fig:burnScanAlpha}).
Thus, to isolate the possible effect on $G_D$ due to the different initial temperatures, we  compare parameter sets with comparable non-diffusive burn fractions $F_{burn}$ (no diffusion).

\begin{figure}[t]
	\center
	\includegraphics[width=\linewidth]{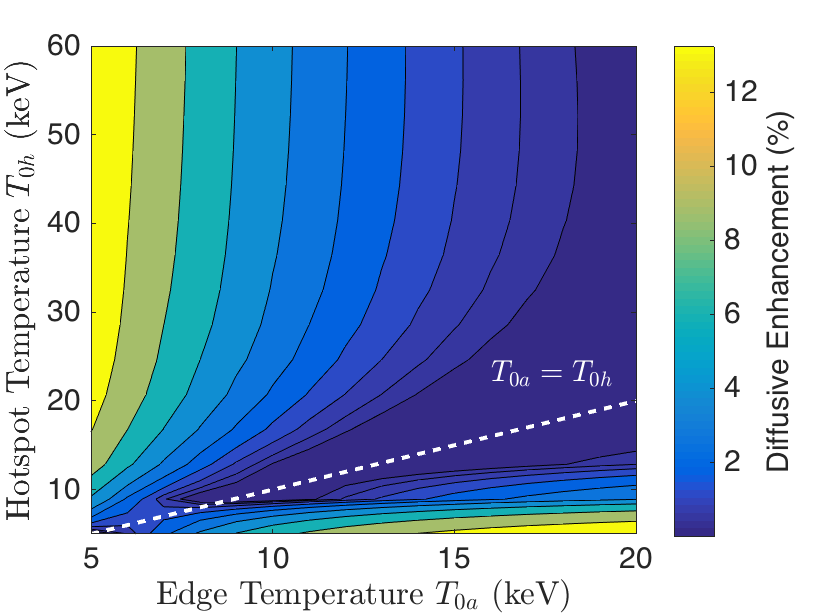}
	\caption{
		Maximum diffusive enhancement $G_D$ over burn rate $S$ as a function of edge temperature $T_{0a}$ and  hot spot  temperature $T_{0h}$,
		with $D_{0a} = 1$.
		At higher $D_{0a}$, the potential enhancements grow larger by a factor of up to $30\%$ [not shown].}
	\label{fig:T0Scan} 
\end{figure}

To accomplish this 
comparison, we vary the factor $S_{0a}$ in Eq.~(\ref{eq:diffModel}), 
leaving constant the other parameters in Eq.~(\ref{eq:diffModel}).
Although such simulations do not necessarily represent physical parameter sets, they isolate the effect of an increased burn 
rate, revealing the maximum potential enhancement from diffusion for a set edge temperature $T_{0a}$ and   hot spot  temperature $T_{0h}$.
The maximal enhancement $G_D^*$ over $S$ is shown in Fig.~\ref{fig:T0Scan} for a variety of parameter sets $(T_{0a},T_{0h})$.
As expected, the maximum possible enhancements occur at the lowest edge temperatures, 
where the fusion reactivity scales strongly with temperature.
Note that the utility of  large hot spot temperatures levels off at around 45 keV,  where the fusion reaction rate $\langle \sigma v \rangle (T)$ becomes flat with temperature.


Although the top left corner of Fig.~\ref{fig:T0Scan} exhibits regimes with enhancements on the order of $12$\%, those regimes are physically hard to access, since the combined constraints of magnetization, high burn fraction, and a short diffusion time necessitate even longer burn times and higher magnetic fields than we are already considering.
Enhancements on the order of 10\%, however, could be accessible for high-field devices with longer, colder burns; for instance, if we take $B_0 = 40$ kT, $T_{0a} = 6$ keV, $n_{0a} = 2 \times  10^{23}$ cm$^{-3}$, $\tau_s = 50 \text{ ns}$, $a = 20$ $\mu$m, and $\alpha = 2$, the simulated enhancement is 10.1\%.

Note that if impurities from the liner are  magnetized, they will  be similarly expelled from the hotspot by the same mechanisms as the ash.
This tendency was observed in mixed-magnetization impurity transport simulations for a non-compressing, wall-confined plasma, where the impurity distribution was found to peak near the point of marginal magnetization
$\Omega_I \tau_{IH} \sim 1$ \cite{vekshtein1975diffusion, ryutov2015characterizing}.
The peaking occurs because near the wall, where the impurities are not magnetized, the thermal forces act oppositely.  
In a cylindrically-compressing plasma with an axial field, the plasma tends to become more magnetized as it compresses, so there will be a tendency toward greater impurity and ash expulsion at later times.

\textbf{Summary and Discussion:} 
%
There is a considerable ongoing campaign for high-yield implosions, but if the fuel burnup is small, then the opportunities offered here do not apply.  
It is only when the fuel burnup is large enough for the ash to poison the fusion reaction that it becomes important to demix the ash from the fuel. 
However, large burnup fraction is necessary for the economical generation of energy in any reactor concept based upon MIF.
The case  here of 25.8\% burnup fraction might be at the low end for economical energy production, but the  4.8\% burnup enhancement is already significant improvement.
Furthermore, these enhancements compare the case of magnetized diffusion to the case of no diffusion.
In fact, the comparison to the case of unmagnetized diffusion could result in an even greater enhancement, since, absent magnetization, highly-charged particles are actually pushed towards regions of higher temperature as a result of thermophoresis effects  \cite{vekshtein1975diffusion}.
The greater enhancement would reflect the avoidance of this deleterious effect in addition to the beneficial demixing effect.


To illustrate these  opportunities for burnup enhancement through magnetized-ion demixing in MIF, we offered a simplified description of magnetized compression,  constrained by both pressure balance and local charge conservation.  
The simplified model neglects radiation, thermal transport, and fuel demixing of D and T. 
Although  DT demixing should saturate at a much lower level than the $\alpha$-fuel demixing, it might somewhat affect the optimal profile steepness, since both fuel and ash demixing increase with profile steepness.
The simplified model also assumes a constant magnetic field, neglecting both the Nernst effect and the collisional exchange of magnetic and thermal pressure due to ion-ion diffusion, which could reduce somewhat the magnetic field in the hotspot.
However, over the time over which the main fusion interactions occur, effects on demixing are likely minor.
Our model  also neglects other effects which could influence demixing, such as plasma rotation \cite{kolmes2018strategies} or turbulent transport from instabilities \cite{weis2014temporal}.
Electron collisions with thermalized ash ions are also neglected, since  thermalized ash  collides primarily with ions.
Initially,  the  slowing down of energetic ash will be dominated by collisions with electrons, affecting the initial distribution of the thermalized ash.  
However, asymmetries in the collisions with electrons would cause energetic alpha particles to be drawn to the colder, denser regions of the plasma even before they thermalize, adding favorably to the demixing of the ash.

The burnup enhancement was optimized over a range of plasma parameters, including the  initial density and temperature profiles, with the key trends identified.
These profiles might be produced by laser-preheating the plasma core.
The large burnup enhancement  occurs in regimes where the fusion reactivity scales strongly with temperature, while satisfying the magnetized transport orderings both for  fuel and thermalized ash and for burn times that accommodate many ion-ion collisions.  
These regimes were shown to be not very far from reactor concepts extrapolated from the present MagLIF-type approach or from magnetized target fusion concepts that feature longer compression times.
They have also been shown to be favorable for other effects, such as magnetic flux conservation \cite{garcia2017mass,garcia2018mass}.

Despite  simplifications, we may conclude that diamagnetic and thermal frictions  can lead to favorable demixing of ash and fuel when the density and temperature profiles are oppositely peaked, as happens naturally in MIF.
Even for burnup fractions of only 25\%, this demixing  increases  fusion power by as much as  5\%.
For economical fusion energy based on  magneto-inertial fusion  approaches, and possibly operating near a point of economic viability, 
a naturally occurring 5\% or greater increase in fusion power  could be highly significant.

\textbf{Acknowledgments:}
The authors thank Elijah Kolmes, Mikhail Mlodik, Seth Davidovits, Steve Slutz, and Paul Schmit for helpful discussions.
This work was supported by NNSA 83228-10966 [Prime No. DOE (NNSA) DE-NA0003764], and by NSF PHY-1506122. 
One author (IEO) also acknowledges the support of the DOE Computational Science Graduate Fellowship (DOE grant number DE-FG02-97ER25308).
\end{document}